\documentclass[a4paper,fleqn,usenatbib]{mnras}

\usepackage[T1]{fontenc}
\usepackage{ae,aecompl}

\usepackage{graphicx}	
\usepackage{amsmath}	
\usepackage{amssymb}	

\usepackage{times}

\usepackage{pdflscape}

\newcommand {\C}{\v Cerenkov }

\newcommand{\nustar}{{\it NuSTAR}}
\newcommand{\swift}{{\it Swift}}
\newcommand{\fermi}{{\it Fermi}}
\newcommand{\suzaku}{{\it Suzaku}}

\title[Hard-TeV BL Lacs with NuSTAR]{The NuSTAR view on Hard-TeV BL Lacs}

\author[L. Costamante et al.]{L. Costamante$^{1,2}$ \thanks{E--mail: luigi.costamante@asi.it}, 
G. Bonnoli$^{2,3,4}$, F. Tavecchio$^{2}$, G. Ghisellini$^{2}$, G. Tagliaferri$^{2}$,
\newauthor{D. Khangulyan$^{5,6,7}$}  \\  
$^{1}$ASI -- Unit\`a Ricerca Scientifica, Via del Politecnico snc, I-00133, Roma, Italy\\
$^{2}$INAF -- Osservatorio Astronomico di Brera, via E. Bianchi 46, I--23807 Merate, Italy \\
$^{3}$Universit\`a degli Studi di Siena, Dipartimento di Scienze Fisiche, della Terra e dell'Ambiente, via Roma 56, I-53100 Siena, Italy \\
$^{4}$INFN -- Sezione di Pisa,  Largo Bruno Pontecorvo 3, I-56127 Pisa, Italy \\
$^{5}$Rikkyo University, Department of Physics, 3-34-1, Nishi-Ikebukuro, Toshima-ku, Tokyo 171-8501, Japan \\
$^{6}$RIKEN iTHEMS, Hirosawa 2-1, Wako, Saitama 351-0198, Japan \\
$^{7}$JAXA, Institute  of  Space  and  Astronautical  Science, 3-1-1  Yoshinodai,  Chuo-ku,  Sagamihara,  Kanagawa  252-5210,  Japan 
}

\date{Accepted 2018 March 31. Received 2018 February 22; in original form 2017 November 11.}

\pubyear{2017}

\begin{document}
\label{firstpage}
\pagerange{\pageref{firstpage}--\pageref{lastpage}}
\maketitle

\begin{abstract}
Hard-TeV BL Lacs are a new type of blazars characterized by a hard intrinsic TeV spectrum,
locating the peak of their gamma-ray emission in the spectral energy distribution (SED) above 2-10 TeV.
Such high energies are problematic for the Compton emission, using a standard one-zone leptonic model. 
We study six examples of this new type of BL Lacs in the hard X-ray band with {\it NuSTAR}. 
Together with simultaneous observations  with the Neil Gehrels {\it Swift} Observatory, 
we fully constrain the peak of the synchrotron emission in their SED,
and test the leptonic synchrotron self-Compton (SSC) model. 
We confirm the extreme nature of 5 objects  also in the synchrotron emission.
We do not find evidence of additional emission components in the hard X-ray band.
We find that a one-zone SSC model can in principle reproduce the extreme properties of both peaks in the SED,
from X-ray  up to TeV energies, but at the cost of  i)  extreme electron energies with very low radiative efficiency,
ii) conditions  heavily out of equipartition   (by 3 to 5 orders of magnitude),
and iii) not accounting for the simultaneous UV data,  which then should 
belong to a different emission component,  possibly the same as the far-IR (WISE) data.
We find evidence of this separation of the UV and X-ray emission in at least two objects. 
In any case, the TeV electrons must not ``see'' the UV or lower-energy photons, 
even if coming from different zones/populations, or the increased radiative cooling would 
steepen the VHE spectrum.
\end{abstract}

\begin{keywords}
BL Lacertae objects: general -- radiation mechanisms: non-thermal -- gamma-rays: galaxies -- X-rays: galaxies
\end{keywords}



\section{Introduction}
In recent years, a new type of blazar has been discovered 
through observations at Very High Energies (VHE, $>$0.1 TeV):  the hard-TeV BL Lacs. 
As any blazar, they are jetted  Active Galactic Nuclei (AGN) with the relativistic jet pointing close to the line of sight.
Their spectral energy distribution (SED) is characterized by two broad peaks, at low and high energy,
commonly explained  as synchrotron and inverse Compton (IC) emission from a population of relativistic electrons in the jet.
Their SED shows the typical properties of high-energy-peaked BL Lacs \citep[HBL,][]{padovanigiommi95}:
an X-ray band fully dominated by the synchrotron emission, a synchrotron peak frequency above the UV band,
and a hard \fermi-LAT spectrum \citep[i.e. photon index $\Gamma_{\rm LAT} \leq$2,][]{3LAC}.
Hereafter, we will refer to a spectrum as hard or steep if 
it is rising or declining with energy in the SED, i.e. if $\Gamma <$ or $> 2$, respectively.

Their intrinsic VHE emission is characterized by 
a hard spectrum ($\Gamma_{\rm VHE}\lesssim$1.5 -- 1.9) 
after correction for the effects of $\gamma$-$\gamma$~ interactions with the  diffuse Extragalactic Background Light 
(EBL),   
even assuming the lowest possible EBL density provided by galaxy counts
\citep[see e.g.][and references therein]{franceschini08,dominguez11,costamante13}.
This locates their Compton peak in the SED  assuredly above 2-10 TeV, the highest Compton-peak energies ever seen in blazars   
\citep[e.g.][]{nature_ebl,hess0347,hess0229,veritas1218,veritas0710}.

This is different from all the other VHE-detected HBL, and in particular those bright in \fermi,
which are characterized by steep intrinsic VHE spectra ($\Gamma_{\rm VHE}>2$)     
and Compton peak energies around $\sim100-200$ GeV. 
The SED of these  ``standard'' HBL can be easily reproduced with a one-zone leptonic emission model 
through the synchrotron self-Compton (SSC) mechanism,  using rather standard parameters.  
A Compton-peak energy above 2-10 TeV becomes instead problematic for standard one-zone leptonic models.  
The decrease of scattering efficiency in the Klein-Nishina regime tends to steepen the gamma-ray spectrum at VHE,
together with the lower energy density of synchrotron seed photons available for scatterings in the Thomson regime, 
as the energy of electrons increases \citep[see e.g.][]{fossati08,tavecchio10}.

Many different alternative scenarios have been proposed: from extremely hard, Maxwellian particle 
distributions \citep{sauge04,lefa11}  to a ``low-energy'' cutoff of the electron distribution 
at very high energies \citep{katarzynski, tavecchio2009};
from internal $\gamma-\gamma$  absorption on a narrow-band radiation field at the source \citep{felix08}
to a separate origin of  the X-ray and TeV emission, the latter coming from kpc-scale jets 
\citep{bottcher08} or as secondary emission 
produced by cascades initiated by ultra-high energy (UHE) protons \citep[e.g.][]{essey1,essey2,prosekin}.
The very low flux or upper limits observed in the GeV band with \fermi-LAT \citep{1LAC,3LAC}
seem now to disfavour internal absorption, at least in these sources.
Likewise, relatively fast variability by a factor a few  observed at VHE 
in 1ES\,1218$+$304 \citep[on a daily timescale,][]{veritas1218var} and 
1ES\,0229$+$200 \citep[on a yearly timescale,][]{veritas0229} 
seem to disfavour both kpc-scale jets (due to the large sizes) and UHE protons.
In the latter case, 
the deflections in the intergalactic magnetic field and the electromagnetic cascades themselves are expected 
to introduce 
long time delays that should smear out such variability, at least up to a few TeV 
\citep[see e.g.][]{prosekin,costamante13}.

In the X-ray band, {\it Beppo}SAX observations have revealed that HBL can reach synchrotron peak frequencies
above a few keV up to and beyond $\sim$100 keV, like Mkn 501 during flares \citep{pian98} and 1ES\,1426+428 \citep{costamante01}.
These states/sources were therefore called  ``{\it extreme synchrotron}'' BL Lacs \citep{costamante01}.
Correspondingly, these hard-TeV BL Lacs could be  called  ``{\it extreme Compton}'' BL Lacs.

Although the two ``extreme'' properties are expected to be coupled,  representing  
the highest-energy end of the blazar sequence \citep{gg98,sequence2},  observationally the relation is not yet clear.
Some objects do appear extreme in both synchrotron and Compton emissions, 
such as 1ES\,0229+200 \citep{ignacio,kaufmann}, 1ES\,0347-121 \citep{hess0347} and  RGB\,J0710$+$591 \citep{veritas0710}.
However, there are cases of objects which are extreme only in Compton but not in synchrotron,
such as  1ES\,1101-232 \citep{gerd1101}, 1ES\,1218$+$304 \citep{veritas1218} and 1ES\,0414$+$009 \citep{hess0414}.
Others are extreme in synchrotron but not in Compton,
such as 1ES\,1426+428 (which is characterized by a steep VHE spectrum  after correction for absorption 
with recent EBL calculations, see e.g. \citealt{dominguez13}).

To reproduce a hard TeV spectrum  most scenarios rely on electron distributions with hard features, 
in the form of ``pile-up'',  narrow-peaked distributions or low-energy cutoffs at high energy. 
If true, such spectral features should become visible also in the synchrotron part of the SED, between the UV and hard X-ray bands, 
because  {\it a)} the synchrotron emission traces directly the energy distribution of the underlying particle population, 
and {\it b)} the expected fluxes (assuming L$_{\rm C}\sim$ L$_{\rm S}$) are comparable to the observed X-ray flux in these sources. 
However, they have not been clearly identified so far. 

Here we report on a set of coordinated observations in the UV to hard-X-ray bands performed on six hard-TeV BL Lacs,
with the satellite \nustar~ \citep{nustar}  and with the X-ray Telescope \citep[XRT,][]{xrt} 
and the Ultra-Violet Optical Telescope \citep[UVOT,][]{uvot} on board \swift.
The goal is to fully characterize the synchrotron hump in the SED and
to identify the synchrotron emission of these putative hard TeV electrons.
This emission could appear in the hard X-ray band beyond 10 keV if due to an  additional population of electrons,
or as a very high X-ray/UV flux ratio, if due to a low-energy cut-off.
In the UV band  the contribution of the host galaxy is generally negligible (for a typical spectrum of elliptical galaxies), 
thus the UV photometry should measure directly the non-thermal synchrotron emission from the jet.
For all sources we also analyzed  the \fermi-LAT data  covering the \nustar~ and \swift~ observations.
Together with the VHE data, these measurements allow us to test the viability of the SSC scenario in such extreme conditions.


\section{Observations and Data Analysis}
The targets were selected on the basis of VHE spectral hardness and highest TeV-peak flux after 
correction for EBL absorption. In addition to archival data on 1ES\,0229+200 ($z$=0.140), the best example of this new class of BL Lacs, 
new observations were performed on 1ES\,0347-121 ($z$=0.188), 1ES\,0414$+$009  ($z$=0.287), RGB\,J0710$+$591 ($z$=0.125), 
1ES\,1101-232 ($z$=0.186) and 1ES\,1218$+$304 ($z$=0.182).
In total, we analyzed  contemporaneous \nustar~ and \swift~  observations of six hard-TeV BL Lacs , 
three of which are characterized historically by hard X-ray spectra, and three by steep X-ray spectra.

The log of the X-ray observations is reported in Table \ref{obs}.
The \swift~ snapshots mostly overlap with the \nustar~ pointings, covering approximately 10\% of the \nustar~ exposure,
with the exception of 1ES\,0414+009 for which the \swift~  observation took place $\sim$18 hours later.
For all sources we checked for variability within and between pointings.  
No relevant variations were observed (flux variations are limited to a few percent), therefore
to improve the statistics on the spectral parameters we considered all the data together, for each source, 
leaving the cross-normalization between instruments free to vary.

\begin{landscape}
\begin{table}
\centering
\caption{Observation logs of the data used in this paper: \nustar~ pointings together with the matching \swift~ observations.
Col. [1]: Object name .
Col. [2 -- 5]: Observation ID, start--stop times and exposure (in seconds) of the \nustar~ observation
Col. [6 -- 9]: Observation ID, start--stop times and exposure (in seconds) of the \swift-XRT data overlapping with (or closest to) the \nustar~ observation. 
Col. [10]: UVOT filter used in the corresponding Obs ID. 
}
\begin{tabular}{lccccccccc}
\hline
\hline
               & \multicolumn{4}{c}{NuSTAR}  & \multicolumn{4}{c}{SWIFT-XRT}  \\
Object         & Obs ID &  START &  END   & Exposure &  Obs ID   &  START &  END   & Exposure  & UVOT            \\
~[1]           &[2]     &[3]     &  [4]   &    [5]   &     [6]   &[7]     &[8]     &   [9]     &[10]   \\ 
\hline 
1ES\,0229+200  & 60002047002  &  2013-10-02T00:06:07 &  2013-10-02T09:31:07  &  13363 &   00080245001  & 2013-10-01T23:40:23 & 2013-10-01T23:52:54 & 394  &  M2   \\ 
               &              &                      &                       &        &   00080245003  & 2013-10-02T00:49:30 & 2013-10-02T18:41:56 & 2359 &  W1   \\
               & 60002047004  &  2013-10-05T23:31:07 &  2013-10-06T10:16:07  &  19784 &   00080245004  & 2013-10-06T00:56:37 & 2013-10-06T09:11:54 & 5209  & W1   \\
               & 60002047006  &  2013-10-10T23:11:07 &  2013-10-11T08:41:07  &  17920 &   00080245005  & 2013-10-10T23:50:32 & 2013-10-10T23:59:56 & 549   & W1    \\
               &              &                      &                       &        &   00080245006  & 2013-10-11T01:32:11 & 2013-10-11T10:58:56 & 5344  & UU    \\
\hline
1ES\,0347-121  & 60101036002  &  2015-09-10T04:51:08 & 2015-09-10T22:01:08   &  32802 &   00081692002  & 2015-09-10T06:02:32 & 2015-09-10T12:31:54 & 1965  & W1    \\
\hline
1ES\,0414+009  & 60101035002  &  2015-11-25T17:01:08 & 2015-11-26T23:01:08   & 34260$^a$      & 00081691001 & 2015-11-27T17:27:16 & 2015-11-27T23:57:55 &  1711 &  W2      \\ 
               &              &                      &                       & 22939$^b$  &           &                   &             &       \\    
\hline
RGB\,J0710+591 & 60101037002  &  2015-09-01T05:51:08  &   2015-09-01T07:46:08  &  3386   & 00081693002  &  2015-09-02T00:03:16  &  2015-09-02T03:37:53  &  4427 & all  \\  
               & 60101037004  &  2015-09-01T12:11:08  &   2015-09-02T01:31:08  &  26365  &           &                   &             &       \\    
               & 60101037006  &  2015-09-02T05:51:08  &   2015-09-02T07:56:08  &  4641   &           &                   &             &       \\    
\hline
1ES\,1101-232  & 60101033002  &  2016-01-12T21:01:08  &   2016-01-14T01:46:08   &  51522   &  00081689001  & 2016-01-12T22:27:16  & 2016-01-12T22:54:54  & 1633  & W1  \\  
               &              &                       &                         &          &  00081689002  & 2016-01-13T00:07:13  & 2016-01-13T02:06:22  & 2854  & UU  \\  
\hline
1ES\,1218+304  & 60101034002  &  2015-11-23T01:06:08  &   2015-11-24T04:11:08   &  49025   &  00081690001  & 2015-11-23T19:39:39  & 2015-11-23T21:26:54  &  1985 & W2  \\
\hline		 
\hline  	 
\multicolumn{8}{l}{$a$: exposure time with nominal aspect solution (A01 files).} \\
\multicolumn{8}{l}{$b$: exposure time with degraded aspect solution (A06 files), see text.} \\
\end{tabular}	 
\label{obs}
\end{table}
\end{landscape}

\subsection{NuSTAR observations}
Data from both focal plane modules (FPMA \& FPMB)  onboard  the \nustar~ satellite  were  
reprocessed  with the {\it NuSTARDAS} software package  (v1.5.1) jointly developed by the ASI Space Science Data Center (SSDC, Italy) 
and the California Institute of Technology (Caltech, USA). Event files were calibrated and cleaned with standard filtering
criteria with the {\it nupipeline} task, using the CALDB version 20150316 and the OPTIMIZED parameter for the exclusion of the SAA passages.
We checked the lightcurves for stability of the background, and used the stricter TENTACLE=yes option to exclude 
intervals with higher/variable background. This was the case only for 1ES\,0229+200. 
We considered only data taken in SCIENCE observing mode, with the exception of 1ES\,0414+009 (see below).

Spectra, ancillary and response files were produced using the {\it nuproducts} task for pointlike sources,
applying corrections for the PSF losses, exposure map and vignetting. 
The FPMA and FPMB spectra were extracted from the cleaned event files using a circle region of 70--80 arcsec radius,
depending on the source intensity, while the background was extracted from nearby circular
regions of 80 arcsec radius, on the same chip of the source (after checking the position angle of the pointing).
All spectra were binned to ensure a minimum of 20 counts per bin.

When different pointings of the same source were performed, separated by few hours to a few days
(like for RGB\,J0710+591 and 1ES\,0229+200),  we checked that there were not important differences. 
To increase the statistics at high energy, we then co-added the spectra for each focal plane module separately,  
using the task {\it addspec} as recommended by \nustar~ science team.  This task combines the source and background PHA files 
as well as the RMF and ARF files.

\subsubsection{Specific procedure for 1ES\,0414+009}
The scientific data corresponding to the SCIENCE observing mode are collected during time intervals in
which an aspect solution from the on-board star tracker located on the X-ray optics bench is available.
When this solution is not available, the aspect reconstruction 
is derived using data from the three star trackers located on the spacecraft bus (SCIENCE\_SC mode).
In this case the accuracy of the sky coordinates degrades to about 2 arcminutes, mainly due to thermal flexing of the 
spacecraft bus star cameras.
Usually the total exposure in this mode is very small compared to the SCIENCE mode (less than 5\%), 
and thus can be filtered out in the standard screening criteria with no significant penalty. 
However, for the observation of 1ES\,0414+009 it amounts  to almost half of the exposure (23 ks out of 57 ks).  
The spectral information in this mode is  fully valid, but it is simply spread out (somewhat unevenly) on a wider area.
Therefore, to recover this additional exposure, we considered the data in this mode as if coming from an extended source.
We extracted the source spectrum from a circle region of radius 120 arcsec, encompassing all the source hotspots,
and the background from a circle region of 80 arcsec on the same chip but as far away from the source region as possible, avoiding the chip borders.
Response and ancillary files were produced with the extended=yes flag, using the default boxsize=20 value 
(i.e. the size in pixel for the subimage boxes).
We then checked the spectral results of the fits in Xspec against those from the optimal SCIENCE mode, finding no significant difference.
We conclude that it is safe to add these spectra to the simultaneous fitting of the total \nustar~ data, 
with floating cross-normalization parameters, in order to improve the statistical error on the spectral parameters.

Later on we also checked the data with the procedure {\it nusplitsc}, newly introduced in NuSTARDAS v1.6.0, which allows
the creation of separate spectra with different combinations of star trackers, confirming the results of our previous approach.

\subsubsection{Upper limits}
The spectral fitting was performed up to energies where the source is detected above 1 sigma (using the command setplot rebin in Xspec).
With the exception of 1ES\,0229+200, detected over the whole \nustar~ band,  the other targets have been detected only up to 30--50 keV.
Since our goal is to constrain the presence of additional emission components in the SED, 
we derived upper limits in the remaining energy range of \nustar~ (up to 79 keV) with the following procedure.
We extracted sky images in the undetected energy band with {\it nuproducts}, and then summed all images together in XIMAGE
(first separately for FPMA and FPMB, and then the total FPMA$+$FPMB images). 
The total image then corresponds to an FPM image with an exposure time equal to the sum of the FPMA and FPMB times,
making full use of the \nustar~ exposure.
On this total image, we checked again the absence of the source detection and
estimated the background countrate level with XIMAGE tools. 
We then extracted the 3-sigma upper limit countrate in a circle region of radius 30 arcsec centered on the target position,
using the command {\it uplimit} in XIMAGE and the Bayesian approach 
(the prior function is set to the prescription described in \citealt{method2}, see XIMAGE user manual). 
We then converted this countrate value to a flux and SED upper limit with Xspec, 
by using a power-law model with photon index fixed to $\Gamma=2.0$, re-normalized to reproduce the measured countrate,
and the response and ancillary (effective area) files appropriate for the 30 arcsec extraction region. 
The upper limit results are reported in Table 2.

\begin{table*}
\centering
\caption{Fit parameters of the \swift-XRT plus \nustar~ data with a log-parabolic model. 
Col. [1]: object name.
Col. [2]: galactic N$_{\rm H}$ used in the fits, in cm$^{-2}$ (see text).
Col. [3]: X-ray band (in keV)  over which the fit is performed.
Col. [4-5]: photon index $\Gamma$ (at 1 keV) and curvature parameter $\beta$ of the log-parabolic fit.
Col. [6]: unabsorbed 2-10 keV flux, in ergs cm$^{-2}$ s$^{-1}$. 
Col. [7]: reduced $\chi^2$ and degrees of freedoms of the fit.
Col. [8]: $\nu F_{\nu}$ Upper Limit in the \nustar~ band where the source is not detected
(namely, in the band $x$-78 keV where $x$ is the upper energy of the detected band, column [4]). See text for details.   
Col. [9]: slope of the spectrum in the soft X-ray band (range 0.3--2 keV, \swift~ data, power-law fit).
Col. [10]: slope of the spectrum in the hard X-ray band (above 5 keV, \nustar~ data, power-law fit).
Col. [11]: cross-normalization of the \swift~ data (vs \nustar~ FPMA).
}
\begin{tabular}{lccccccccccc}
\hline
\hline
Name      &  N$_{\rm H,gal}$   &  Band   & $\Gamma$ & $\beta$ & $F_{[2-10]}$ &  $\chi^2_{r}$ (d.o.f.)  &  UL & $\Gamma_{0.3-2}$ & $\Gamma_{5-50}$ & K$_{\rm XRT}$   \\
~[1]      &  [2]  &[3]                 &[4]           &[5]       &[6]      &[7]              &[8]                     & [9] &  [10]            &    [11]         \\%
\hline 
1ES\,0229+200  &   9.20 e20 & 0.3--78 & $1.49\pm0.04$  & $0.27\pm0.02$  &   1.95 e-11    &  1.027 (279)  &   --      &  $1.50\pm0.05$   &  $2.03\pm0.02$ &  0.87  \\
1ES\,0347-121  &   3.60 e20 & 0.3--40 & $1.93\pm0.12$  & $0.25\pm0.08$  &   5.67 e-12    &  0.986 (\;99) & 4.12 e-12  &  $1.96\pm0.18$   &  $2.42\pm0.08$ &  0.93 \\	 
1ES\,0414+009  &   9.15 e20 & 0.3--30 & $2.30\pm0.11$  & $0.29\pm0.07$  &   7.53 e-12    &  0.997 (180)  & 1.53 e-12  &  $2.29\pm0.15$   &  $2.80\pm0.07$ &  0.98  \\	  
RGB\,J0710+591 &   4.45 e20 & 0.3--50 & $1.61\pm0.06$  & $0.35\pm0.04$  &   1.71 e-11    &  1.081 (191)  & 1.19 e-11  &  $1.60\pm0.08$   &  $2.34\pm0.04$ &  1.00 \\
1ES\,1101-232  &   5.76 e20 & 0.3--45 & $1.90\pm0.05$  & $0.34\pm0.03$  &   2.57 e-11    &  1.016 (253)  & 6.37 e-12  &  $1.86\pm0.07$   &  $2.56\pm0.03$  & 0.93  \\  
1ES\,1218+304  &   1.94 e20 & 0.3--40 & $1.93\pm0.07$  & $0.36\pm0.05$  &   1.10 e-11    &  0.908 (140)  & 3.63 e-12  &  $1.95\pm0.09$   &  $2.64\pm0.05$  & 1.13  \\
\hline		 
\hline  	 
\end{tabular}	 
\label{xfits}
\end{table*}

\begin{table*}
\centering
\caption{UVOT $V$, $B$, $U$, $W1$, $M2$, $W2$ observed Vega magnitudes for the pointed objects.
Photometry performed on the summed images from all exposures. 
Magnitudes not corrected for Galactic extinction. The extinction values used to de-redden the fluxes for the SED
are reported in the last two columns, as total absorption A$_{\rm B}$ (in magnitudes) and E(B-V)
(from NED, see Sect. 2.2).   
}
\begin{tabular}{lcccccccc}
\hline
\hline
Name           &   $V$           &  $B$             &  $U$            &      $W1$              &       $M2$       &  $W2$           & A$_{\rm B}$  & E(B-V)   \\        
\hline 
1ES\,0229+200  &    --             &      --            & $17.70\pm0.03$  &  $17.96\pm0.03$  &  $18.16\pm0.09$  &     --          &  0.493  & 0.120   \\
1ES\,0347-121  &    --             &      --            &      --        &  $17.81\pm0.04$  &       --        &     --          &  0.167 &  0.040   \\
1ES\,0414+009  & $16.53\pm0.09$     &  $16.84\pm0.06$     & $15.86\pm0.05$  &  $15.81\pm0.05$  &  $15.94\pm0.03$  &  $15.93\pm0.03$  &  0.465 &  0.114  \\
RGB\,J0710+591 & $16.52\pm0.05$     &  $17.38\pm0.04$     & $16.70\pm0.04$  &  $16.60\pm0.04$  &  $16.58\pm0.04$  &  $16.74\pm0.03$  &  0.139 &  0.034  \\
1ES\,1101-232  &    --             &      --            & $16.64\pm0.03$  &  $16.43\pm0.03$  &       --        &      --         &  0.221 &  0.054  \\
1ES\,1218+304  &    --             &      --            &      --        &       --        &      --         &  $15.66\pm0.02$  &  0.074 &  0.018  \\
\hline		 
\hline  	 
\end{tabular}	 
\label{uvot}
\end{table*}

\subsection{Swift observations}
All XRT data were re-processed with standard procedures using the FTOOLS task {\it xrtpipeline}. Only
data taken in photon counting (PC) mode were considered, representing the bulk of the \swift~ exposure.

Source events were extracted in the 0.3--10 keV range within a circle of radius  20 pixels ($\sim 47\arcsec$), while
background events were extracted from circular and annular regions around the target, free of other sources.
When multiple datasets were available, after checking for variability, 
event files were merged together in XSELECT  
and the total energy spectrum  was extracted from the summed cleaned event file following the same procedure.
The ancillary response files were generated with the {\it xrtmkarf} task, applying corrections for the PSF losses 
and CCD defects using the cumulative exposure map.  
When the source countrate was above 0.5 cts/s, spectra were checked for pile-up problems, 
and if necessary the central region of the PSF was excised in the spectral extraction.  
This was the case for RGB\,J0710+591 (inner 3 arcsec radius),  1ES\,1101-232 (7 arcsec) and  1ES\,1218+304 (3 arcsec).   
The source spectra were binned to ensure a minimum of 20 counts per bin.

UVOT  observations were performed with the UV filter of the day (see Table \ref{obs}).
Photometry of the source was performed using the UVOT software in the HEASOFT 6.18 package.
Counts were extracted from the standard aperture of 5 arcsec radius for all single exposures and all filters, 
while the background was carefully estimated from different positions more than 27 arcsec  away from the source,
and avoiding other sources in the field of view. 
Count rates were then converted to fluxes using the standard  zero points \citep{poole}.   
After checking for variability, all the frames for each filter were summed and a total photometry performed. 

For the SED, the fluxes were de-reddened using the mean Galactic interstellar extinction curve  from \citet{fitzpatrick},
using the values of A$_{\rm B}$ and E(B-V) taken from NED, and reported in Table \ref{uvot}.
These are based on the \citet{schlafy} recalibration  of the \citet{schlegel} infrared-based dust map.
For combined UVOT-XRT fits in Xspec, the {\it uvot2pha} tool was used, together with the canned 
response matrices available in the CALDB.

\begin{figure*}
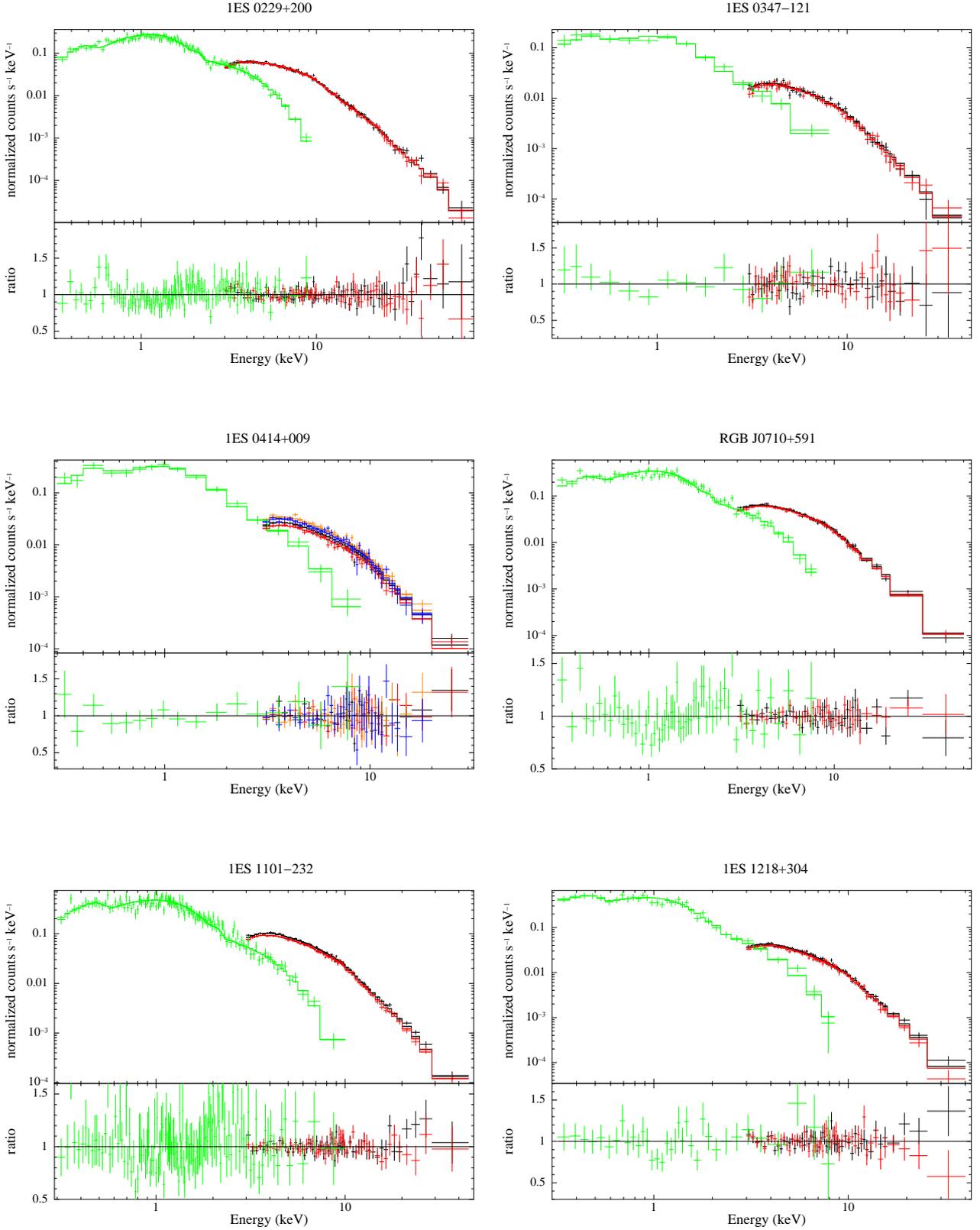

\includegraphics[angle=-90,origin=c,width=8.5cm]{0229xfit.ps}
\includegraphics[angle=-90,origin=c,width=8.5cm]{0347xfit.ps} \\
\includegraphics[angle=-90,origin=c,width=8.5cm]{0414xfit.ps}
\includegraphics[angle=-90,origin=c,width=8.5cm]{0710xfitb.ps} \\
\includegraphics[angle=-90,origin=c,width=8.5cm]{1101xfit.ps}  
\includegraphics[angle=-90,origin=c,width=8.5cm]{1218xfit.ps}  
\caption{X-ray spectra of the six hard-TeV BL Lacs as observed by {\it Swift}/XRT and {\it NuSTAR},
together with the log-parabolic best-fit model and corresponding ratio data/model.
{\it NuSTAR} data are in black and dark grey (black and red in the electronic version), 
while {\it Swift}/XRT data are in light grey (green).
For 1ES\,0414+009, in addition, the {\it NuSTAR} data of FPMA and FPMB taken with degraded aspect solution (A06) 
are shown in grey, with the same binning (blue and orange in the electronic version).  }
\label{xfigs}
\end{figure*} 

\begin{table*}
\centering
\caption{Parameters useful for the SED properties. 
Col. [1]: object name.
Col. [2]: redshift.
Col. [3-4]: energy (in keV) and $\nu F_{\nu}$ flux (in erg cm$^{-2}$ s$^{-1}$)  of the synchrotron peak in the SED, 
as resulting from the log-parabolic X-ray fit (Table \ref{xfits}).
Col. [5]: photon index of the VHE spectrum, after correction for EBL absorption according to the EBL calculations in Franceschini et al. 2008. 
Statistical uncertainties are in the range $\pm0.2-0.4$. Data and values from Aharonian et al. (2006, 2007a, 2007b, 2007c),  Abramowski et al. 2012 and Acciari et al. (2009, 2010a).
Col. [6-7]: approximate lower limits for  the energy (in TeV) and $\nu F_{\nu}$ flux (in erg cm$^{-2}$ s$^{-1}$) of the Compton peak in the SED,
assuming a power-law VHE spectrum.  Namely, upper energy and correspondig flux at the end of the detected VHE band, after correction for EBL asborption.
Col. [8]: slope of the UV to X-ray emission in \swift: namely, photon index of a power-law fit in Xspec of the UVOT datapoints with the XRT 0.3-1 keV data,
accounting for galactic extinction in the optical-uv (model {\it Redden94}) and X-ray bands ({\it wabs}).  
}
\begin{tabular}{lccccccc}
\hline
\hline
Name      &  $z$  &  E$_{peak}$(synch)   &  $\nu F_{\nu,peak}$(synch) &  $\Gamma_{\rm VHE}$ &  E$_{peak}$(Compt) &  $\nu F_{\nu,peak}$(Compt) &  $\Gamma_{\rm UV-X}$  \\
~[1]      &  [2]  &   [3]                &       [4]             &        [5]          &        [6]         &        [7]            &       [8]          \\%
\hline 
1ES\,0229+200  & 0.140 &   $9.1\pm0.7$  &   1.31 e-11   &   1.5    &  $>12$ TeV  & $>2$ e-11   &    1.7 \\ 
1ES\,0347-121  & 0.188 &   $1.4\pm0.6$  &   4.17 e-12   &   1.8    &  $>3$ TeV	 & $>8$ e-12   &    1.8 \\ 
1ES\,0414+009  & 0.287 &   $0.3\pm0.2$  &   1.13 e-11   &   1.85   &  $>2$ TeV	 & $>3$ e-12   &    2.0 \\ 
RGB\,J0710+591 & 0.125 &   $3.5\pm0.2$  &   1.11 e-11   &   1.85   &  $>4$ TeV	 & $>6$ e-12   &    1.8 \\ 
1ES\,1101-232  & 0.186 &   $1.4\pm0.2$  &   1.98 e-11   &   1.7    &  $>4$ TeV	 & $>1$ e-11   &    1.7 \\ 
1ES\,1218+304  & 0.182 &   $1.3\pm0.3$  &   8.97 e-12   &   1.9    &  $>2$ TeV   & $>2$ e-11   &    1.9 \\ 
\hline		 
\hline  	 
\end{tabular}	 
\label{peaks}
\end{table*}

\subsection{X-ray spectral analysis}
For each source, simultaneous fits of the XRT and \nustar~ spectra were performed using the XSPEC package,
using the C-stat statistics adapted for background-subtracted data.

To account for cross-calibration uncertainties between the three telescopes (two \nustar~ and one \swift),
a multiplicative constant factor has been included in the spectral model,
kept fixed to 1 for \nustar/FPMA and free to vary for FPMB and XRT.  In the case of FPMB the difference is of the order of 5\%,
consistent with the expected cross--calibration uncertainty for the instruments \citep{madsen}.

For XRT the difference is of the order of 0-15\%, which is slightly larger than the typical inter-calibration uncertainties ($\lesssim10\%$).
This can be explained with the different exposure times and not fully overlapping epochs, allowing for  residual small flux variations.
We checked the robustness of the spectral fits also using the cross-calibration obtained by fitting the data 
in a strictly common energy band (3-9 keV) with a power-law model, finding fully consistent results.    
For 1ES\,0414+009, the FPMA and FPMB datasets taken with degraded aspect solution are included as well, with free
constant factors. The differences are again within 5\%.

The spectra were fitted with both a single power-law and log-parabolic model, 
with the ISM absorption model {\it wabs} and hydrogen--equivalent column density N$_{\rm H}$ fixed at the Galactic values, 
as given by the HEASARC tool {\it w3nh} using the LAB survey maps \citep{kalberla}.

The only exception is 1ES\,0229+200, for which we preferred to use the Dickey \& Lockman \citep[DL,][]{DL}  value of $9.2\times 10^{20}$ cm$^{-2}$
(vs $8\times 10^{20}$), as a more accurate estimate for the intervening absorption.
The reason is that a)  the most recent relation of column density vs optical extinction 
\citep[N$_{\rm H}$=$8.3\times 10^{21}$ E(B-V),][]{liszt}     
gives N$_{\rm H}$ values of $\sim1\times 10^{21}$,
closer to the DL estimate; and
b) fits with free N$_{\rm H}$ 
tend to prefer higher N$_{\rm H}$ values of $\sim1-1.1\pm0.2 \times 10^{21}$ cm$^{-2}$ than harder power-law models,
even using the \swift~ data alone.  Regardless, the Galactic absorption towards 1ES\,0229+200
should be considered with a systematic uncertainty in N$_{\rm H}$ between 0.8 and 1.1 $\times 10^{21}$ cm$^{-2}$.

For all objects, a single power-law model does not provide a good fit, showing clear evidence of curvature in the residuals.
A log-parabolic model provides a good description of the data in the 0.3--79 keV energy band,
and it allows to estimate the error on the position of the synchrotron peak energy (defined as the energy where $\Gamma_{\rm X}=2.0$) and 
flux \citep[see e.g.][]{tramacere}.     
The results of the fits are presented in Table \ref{xfits}, together with the local slopes of the spectrum in the soft (0.3-2 keV) 
and hard (5-50 keV) X-ray bands.  The spectral fits and residuals of the best fit models are shown in Figure \ref{xfigs}.

\subsection{Fermi-LAT data}
We analyzed  the data from the LAT detector \citep{lat} onboard the \fermi~ satellite 
using the latest Pass 8 data version  and the public Fermi Science Tools version v10r0p5.
Given the very low LAT countrate for these objects,  we chose to integrate the LAT data over the last 4 years,
from 2013 January 1 to 2016 December 20 (MET 378691203, 503926885). This time interval encompasses
all the \nustar~ pointings of our targets (which occurred between 2013 and 2016), 
and provides an homogeneous dataset and time interval over which all targets are detected.
To avoid problems with the strong background at low energies, we selected SOURCE class events in the 0.3-300 GeV range
with {\it gtselect}.
Gamma-ray events were selected from a Region of Interest (ROI) of 15\degr using
standard quality criteria, as recommended by the Fermi Science Support Center (FSSC).
The instrument response functions \verb+P8R2_SOURCE_V6+ were used. 
The Galactic and isotropic diffuse emission were accounted with the models 
\verb+gll_iem_v06.fits+  and  \verb+iso_P8R2_SOURCE_V6_v06.txt+, respectively. 

\begin{table}
\centering
\caption{Parameters of the fits to the Fermi-LAT data with a power-law model, in the 0.3-300 GeV band.
For all sources, the LAT integration time is  from 01/01/2013 to 20/12/2016.
Col. [1]: object name.
Col. [2]: test statistics \citep{mattox}.
Col. [3]: integrated photon flux 1-100 GeV, in cm$^{-2}$ s$^{-1}$ (for comparison with the 3FGL fluxes).
Col. [4]: photon index of the LAT spectrum, in the 0.3-300 GeV band.
}
\begin{tabular}{lccc}  
\hline
\hline
Name           &   TS      &  Flux (1-100 GeV)    &  $\Gamma_{\rm LAT}$   \\
~[1]           &   [2]     &       [3]            &       [4]            \\%
\hline 
1ES\,0229+200  &   \enspace56.7   &  $2.55\pm0.77$ e-10  &  $1.49\pm0.18$    \\
1ES\,0347-121  &   \enspace48.8   &  $2.56\pm0.62$ e-10  &  $1.64\pm0.16$    \\
1ES\,0414+009  &          125.3   &  $7.00\pm0.95$ e-10  &  $1.95\pm0.11$     \\
RGB\,J0710+591 &   \enspace96.6   &  $3.44\pm0.62$ e-10  &  $1.67\pm0.12$     \\
1ES\,1101-232  &   \enspace63.3   &  $3.47\pm0.78$ e-10  &  $1.65\pm0.15$     \\
1ES\,1218+304  &          972.1   &  $2.67\pm0.17$ e-09  &  $1.72\pm0.04$     \\
\hline		 
\hline  	 
\end{tabular}	 
\label{fermi}
\end{table}

Because our time interval does not overlap with the epoch used for the 3FGL catalog,
we performed the likelihood analysis using {\it gtlike} in two steps. In the first step, using the tool
\verb+make3FGLxml.py+  we included in the XML model file
all sources in the 3FGL catalog, with free parameters for those within an 8-degree radius from the target or
with free normalization for any source in the ROI flagged as variable in the 3FGL.
We then checked the residual maps --in both counts and Test Statistic \citep[TS,][]{mattox}-- 
for new sources or unmodeled variations of known sources, and changed the XML file accordingly,
identifying and adding sources if TS$>20$ and dropping all 3FGL sources with a TS$<1$ in the time interval considered here. 
We  then performed a second likelihood analysis, using the new optimized XML model file.
The analysis was performed using a binned likelihood with 0.1 deg bins and 10 bins for decade in energy, 
with the NEWMINUIT optimizer.

All our targets are detected above 5$\sigma$  in the considered interval. 
The spectra are well fitted with a single power-law model in the 0.3-300 GeV range.
Fit parameters are reported in Table \ref{fermi}.
For the SED, the LAT data points were obtained performing a likelihood analysis in each single energy bin,
using as model the XML file of the global fit with all parameters fixed to the best-fit values. 
Only the normalization of the target and of the two backgrounds was left free.  
A binned or unbinned likelihood was used if the total number of counts in the bin was higher or lower than 1000, respectively.
A Bayesian upper limit is calculated if in that bin the target has a TS$<9$ or {\it npred}$<2$.

\section{Results}
All sources  show significant curvature in their broad-band X-ray spectrum,
which pins down the location of their synchrotron peak in the SED.

The peak energies  calculated from the log-parabolic model are reported in Table \ref{peaks}, 
together with other SED properties constrained by the data:
the slope of the spectrum between the  UV and soft X-ray (0.3-1 keV) bands,  
the photon index of the VHE spectrum after correction for EBL absorption  \citep[with the model by][]{franceschini08}
and the corresponding lower limits on the energy  and flux of the Compton peak in the SED. 

Five objects present the synchrotron peak directly inside the observed X-ray band, 
revealing  their extreme-synchrotron nature. 
The peak energy is around 10 keV for 1ES\,0229+200, and in the range 1--4 keV for the other four objects.
1ES\,0229+200 therefore confirms its most extreme character in the synchrotron as well as Compton emission.
It is the only object in our sample detected over the entire \nustar~ band up to 79 keV, and it is also characterized by 
the hardest slope in the soft X-ray band   ($\Gamma_X \simeq1.5$, with a systematic uncertainty of $\pm0.1$ due 
to the uncertainty on the Galactic N$_{\rm H}$ values).

1ES\,0414+009 is instead the only object in our sample showing  more ``standard'' HBL properties. 
Its X-ray spectrum is steep ($\Gamma_X >2$) over the whole band,
locating the synchrotron peak energy below or around 0.3 keV (from the curvature of the spectrum).

In all sources there is no evidence of spectral hardening at the highest X-ray energies (see Fig. \ref{xfigs}).
The \nustar~ data above 10 keV do not show any significant excess above the spectrum determined in the 0.3-10 keV 
range. The synchrotron emission of the hard-TeV electrons  does not seem to provide any significant contribution
in the observed hard X-ray band. Therefore any additional emission component in the SED 
is constrained to appear well above 100 keV. 

The UVOT data are generally consistent with the soft X-ray spectrum belonging to the same emission component in the SED.
In other words, the slope measured in the soft X-ray band is equal or steeper than the power-law slope
connecting the UV to the  soft X-ray fluxes ($\Gamma_{\rm 0.3-2 keV}\gtrsim \Gamma_{UV-X}$).

However, this is not the case in two objects, namely 1ES\,0229+200 and RGB\,J0710+591.
Their soft X-ray spectrum is harder than the UV-to-X-ray slope by about $\Delta\Gamma\sim 0.2$,  well 
above the statistical uncertainty.
The UV data therefore remain significantly  above the power-law extrapolation of the soft X-ray spectrum towards lower energies.
This ``UV excess'' can either be due to a different synchrotron component, from a different electron population,
or could be explained by unaccounted thermal emission from the host galaxy, for example because of a burst of massive star formation
caused by a recent merger. However, the additional thermal UV flux required in the latter case would be large, 
one to two orders of magnitude above the template of a giant elliptical galaxy \citep[][see following section]{silva}, 
and therefore we consider this explanation less feasable.

In the gamma-ray band, 
the \fermi-LAT spectrum match surprisingly well with the VHE spectrum in both flux and slope, 
after correction for EBL absorption (see Table \ref{peaks} and \ref{fermi}).
The photon indices are similar, in the 1.5--1.8 range. This is the case also for 
the hardest object,  1ES\,0229+200, which shows the same photon index $\Gamma\approx1.5$ both in the \fermi-LAT and VHE bands.  
This points toward a Compton peak in the SED  well in excess of 10 TeV.

The only exception is represented by 1ES\,0414+009. The flat $\Gamma_{\rm LAT}\simeq 2$ \fermi-LAT spectrum  
does not seem to belong to the same SED emission component of the VHE data (see Fig. \ref{seds}).
The VHE spectrum appears brighter and harder, though still within the statistical uncertainty.
This is one of the weakest BL Lacs detected so far in the VHE band ($\sim0.6$\% of the Crab Nebula flux). 
However, the LAT spectrum extracted at the beginning of the \fermi~ mission and 
partly overlapping with the H.E.S.S. observations \citep[years 2008-2010, see][]{hess0414}  
is characterized by 
a slightly harder photon index and higher flux, more in line with the intrinsic VHE spectrum \citep{hess0414}. 
We confirm this using Pass 8 data ($\Gamma_{\rm LAT}=1.77\pm0.14$).
We conclude therefore that 1ES\,0414+009 has likely changed its gamma-ray SED properties between the two epochs.

\section{SED modeling}
\begin{figure*}
\vspace{1.3cm}
\includegraphics[width=8.0cm]{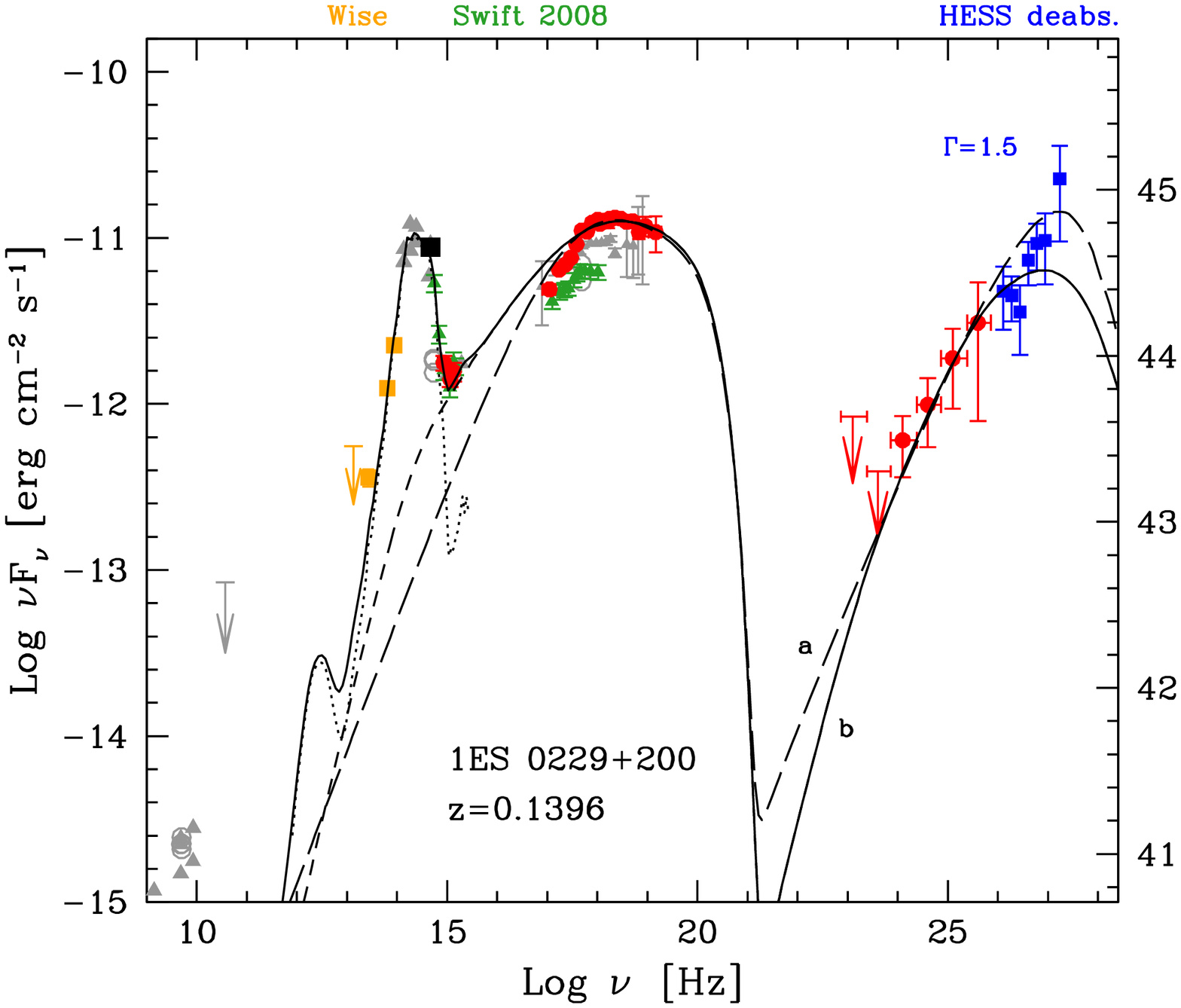}\hspace{8mm}
\includegraphics[width=8.0cm]{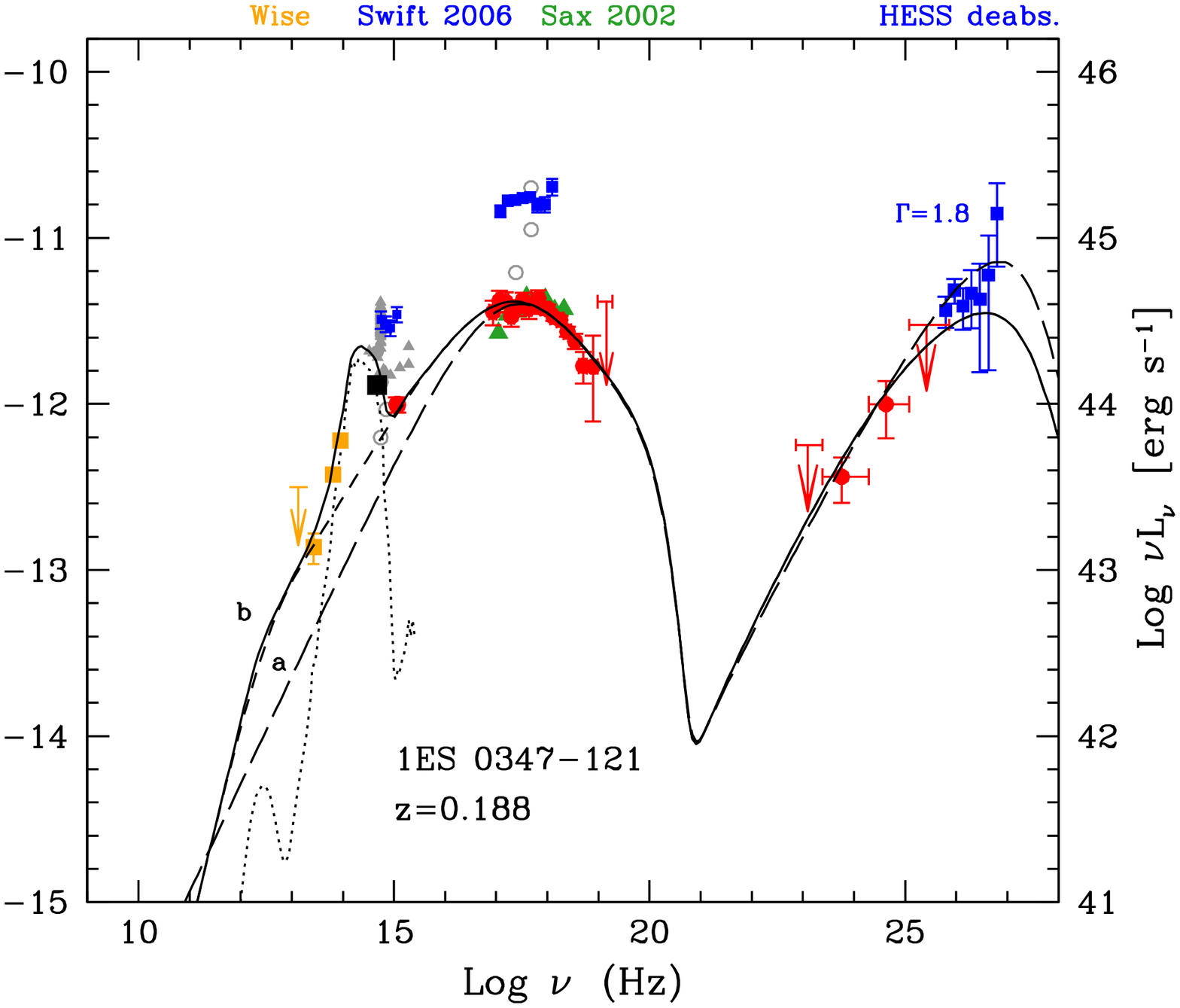} \vspace{1.6cm}  \\
\includegraphics[width=8.0cm]{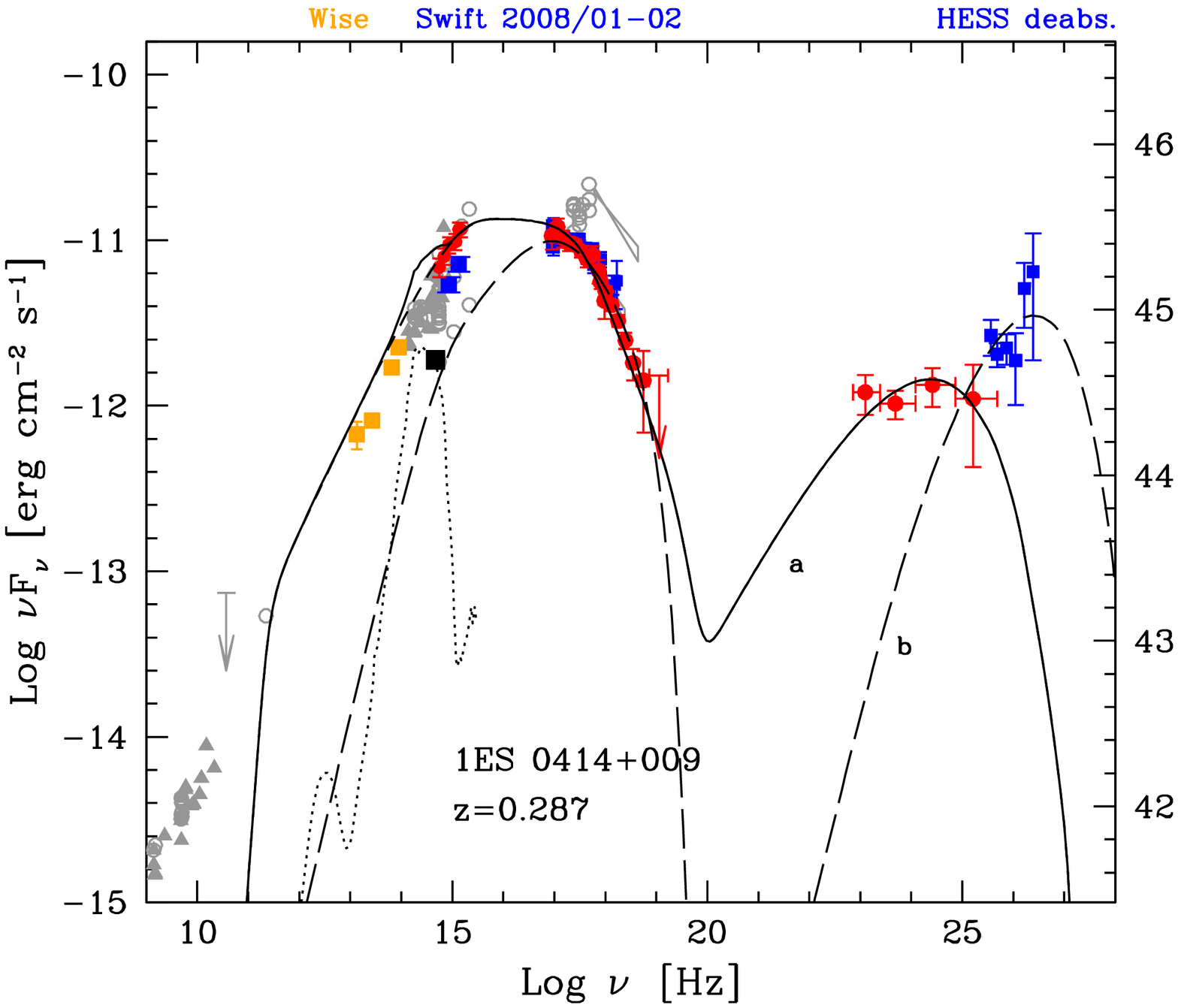}\hspace{8mm}
\includegraphics[width=8.0cm]{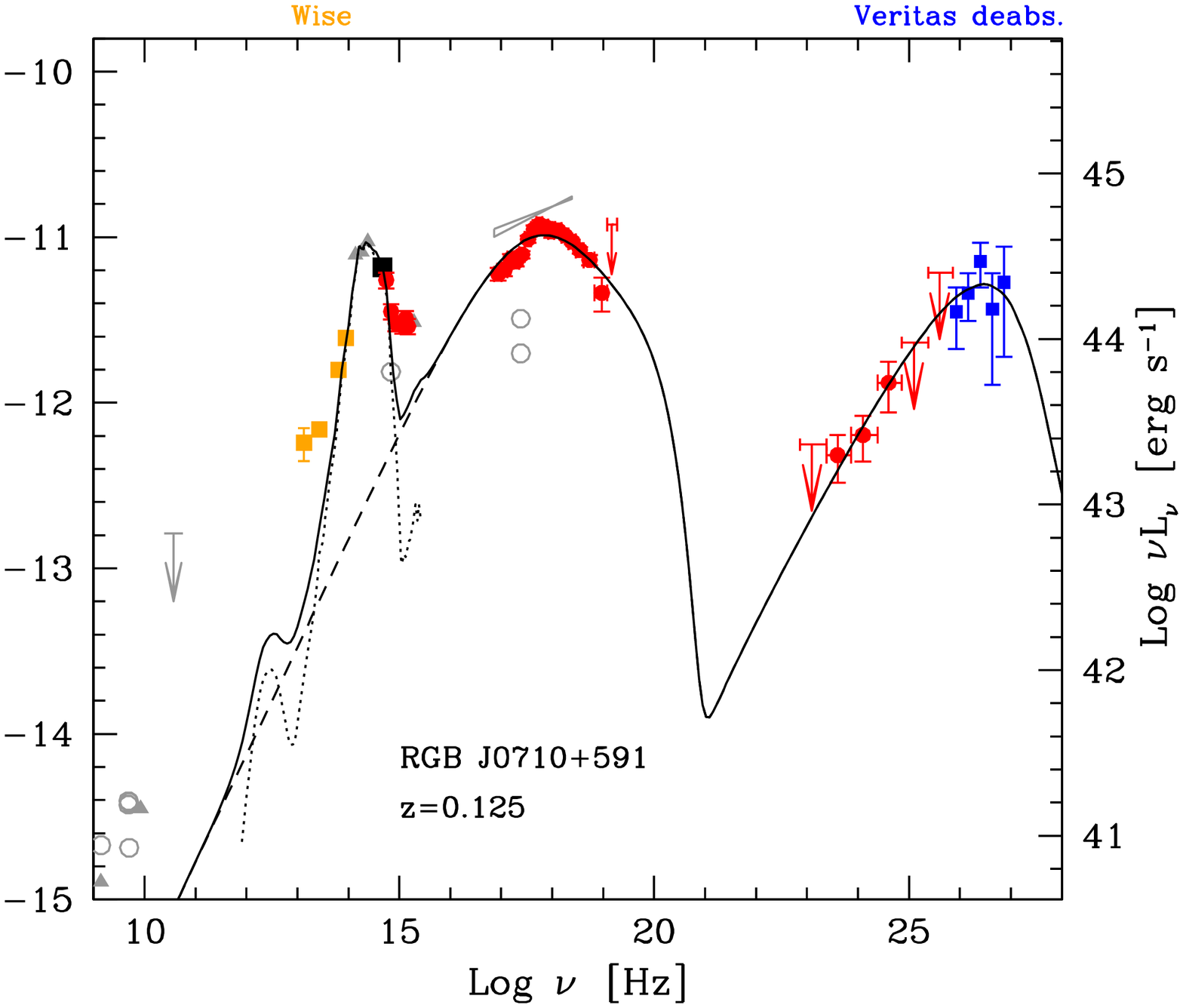} \vspace{1.6cm}  \\
\includegraphics[width=8.0cm]{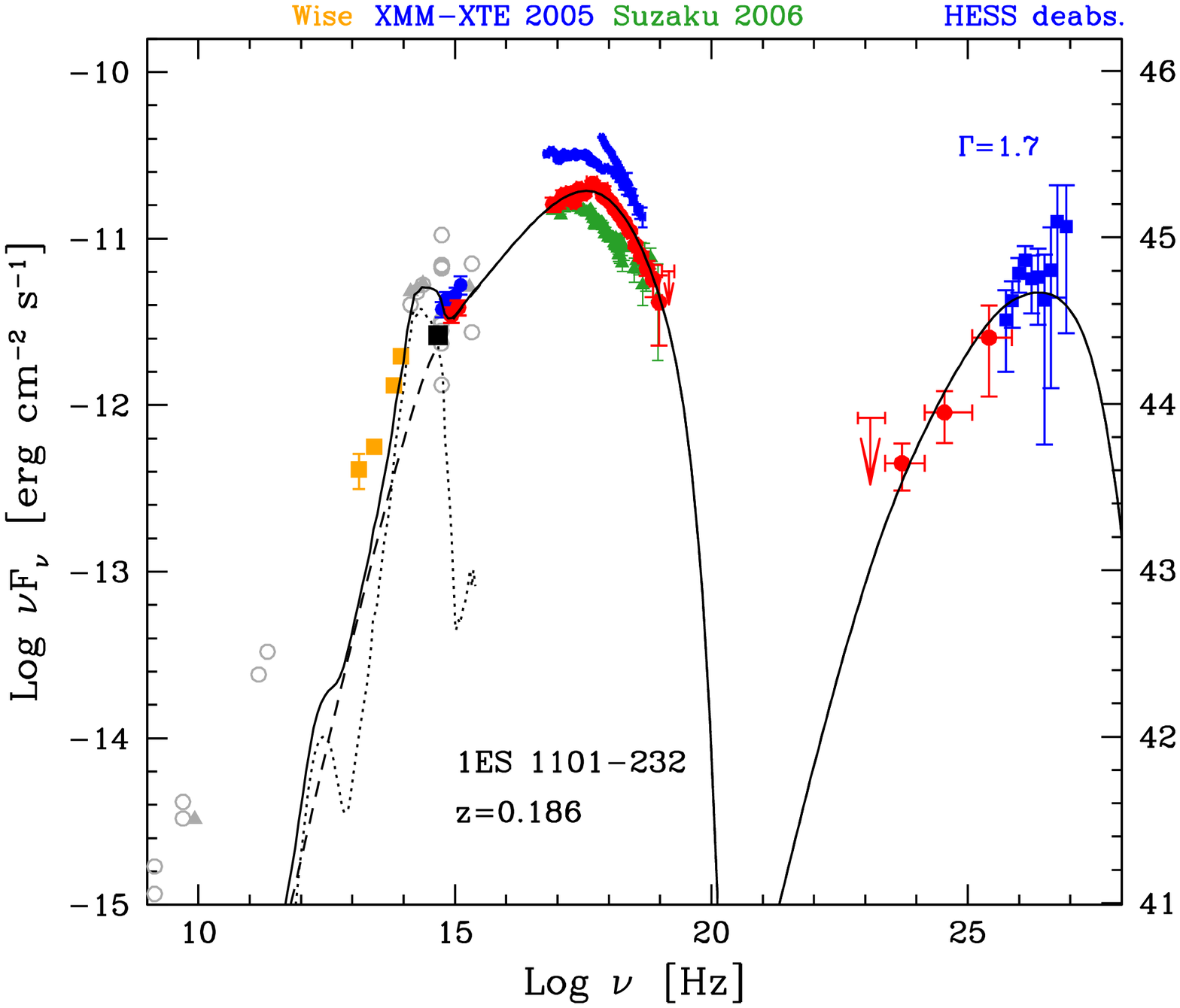} \hspace{8mm}
\includegraphics[width=8.0cm]{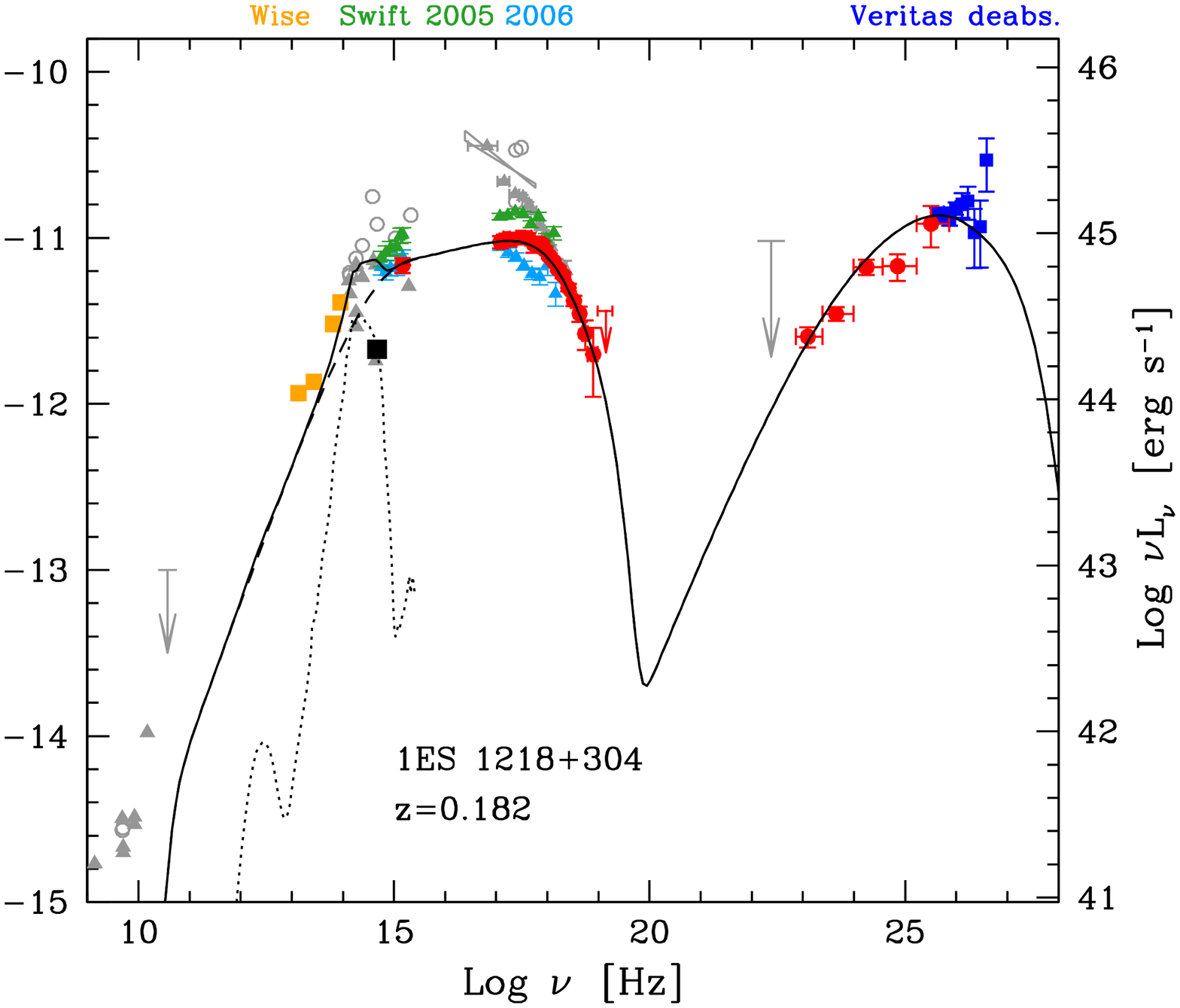}  
\caption{Overall SEDs of the six hard-TeV BL Lacs considered here. The new data presented in this paper 
(\swift, \nustar~ and \fermi~ observations)
are shown as filled circles (red in the electronic version). Historical data taken close to the same epoch of the VHE data 
are shown as filled squares (blue).
Other non-contemporaneous archival data are reported in grey as filled triangles or open circles 
(in color in the electronic version, if recent \swift~ or \suzaku~ data, as labelled).
The VHE data are corrected for EBL absorption following \citet{franceschini08}.
Solid lines show the sum of the theoretical SSC model (short-dashed line) and host-galaxy emission (dotted line).  
Long and short dashed lines show the SSC modeling of the SED for one or two different sets of parameters
reported in Table \ref{param}, as labelled.
The SED of the host galaxy is obtained using the template of a giant elliptical galaxy from \citet{silva}
renormalized to the magnitude of the resolved host galaxy \citep[][black full squares]{scarpa2000, falomo2000}.
For each SED, the vertical axis on the right shows the Luminosity values in Log($\nu L_{\nu}$) scale (erg s$^{-1}$). 
}
\label{seds}
\end{figure*}

We assembled the SEDs of these BL Lacs complementing   our data  
with archival data from the NASA/IPAC Extragalactic Database (NED), the ASI Space Science Data Center (SSDC) and  
the Wide-field  Infrared  Survey Explorer  (WISE\footnote{Data retrieved from the {\it WISE} All--Sky Source Catalog: 
\url{http://irsa.ipac.caltech.edu/.}}) satellite \citep{wise}.
To account for the presence of the host galaxy --which are typically ellipticals in blazars-- 
we adopted the SED template of a giant elliptical galaxy  from \citet{silva}, 
renormalized to match the magnitude of the resolved host galaxy \citep{scarpa2000, falomo2000}. 
We corrected the VHE data for the effects of EBL absorption with the model by \citet{franceschini08},
which agrees well with all present limits and the other recent calculations between 
UV-Optical and mid infrared wavelengths.
The EBL spectrum in its direct stellar component is now well constrained within a narrow
band close to the lower limits given by galaxy counts
\citep[see e.g.][]{nature_ebl,lat_ebl,dominguez11,costamante13}.
Figure \ref{seds} shows the overall SEDs.

We tested the one-zone leptonic SSC scenario
using the emission model fully described in \citet{tavecchio03}.
In summary,  the emitting region is assumed to be a sphere  of radius $R$  entangled with  uniform magnetic field $B$. 
The distribution of  relativistic electrons is assumed isotropic and follows a smooth broken power law energy 
spectrum with normalization $K$ and indices $n_1$ from $\gamma_{\rm min}$ to $\gamma_{\rm b}$ and $n_2$ above 
the break up to $\gamma_{\rm max}$.   These electrons emit through synchrotron and SSC mechanisms. 
The SSC emission is calculated assuming the full Klein-Nishina cross section, which is important in these objects given 
the high electron energies.
Bulk-motion  relativistic amplification of the emission is described by the Doppler factor $\delta$. 
In total, nine parameters ($R$, $B$, $K$, $n_1$, $n_2$, $\gamma _{\rm min}$, $\gamma _{\rm b}$, $\gamma _{\rm max}$, $\delta$) 
fully specify the model.

The theoretical SEDs calculated with this model are plotted in Figure \ref{seds}, with the corresponding 
parameters reported in Table \ref{param}.
To match the gamma-ray data and achieve a Compton peak in the multi-TeV range with sufficient luminosity, 
the SSC modeling requires  
extreme energies of the peak electrons,  close to 10$^6$ $m_{\rm e} c^2$,  
extremely low magnetic fields, at the milliGauss level  
--implying conditions  out of equipartition by 3 to 5 orders of magnitude-- 
and  a low density of soft photons in the emitting region.
The latter condition is essential in order to avoid  efficient cooling of the TeV electrons in the Thomson regime,  
which would lead to much softer gamma-ray spectra in the VHE range. 
In a one-zone leptonic scenario, this requires  electron distributions with hard low-energy spectra --or low-energy cutoffs--
below the X-ray band.

This scenario seems indeed corroborated by our data in the two objects with an ``UV excess'' (1ES\,0229+200 and RGB\,J0710+591).
In these objects the UV flux remains above the extrapolation of the soft X-ray power-law spectrum to lower energies,
indicating that they might belong to a different emission component.   
This is also suggested by the WISE fluxes at the lowest-frequencies, 
which  stay above the host-galaxy emission 
and can be connected with the UV fluxes as a single power-law emission component (see Fig. \ref{seds}).
The important point is that these two components must not ``see'' each other:
allowing the X-ray electrons to upscatter the observed IR--UV synchrotron photons, whatever their origin, 
would make impossible to reproduce the gamma-ray data, because of the much softer spectra as a result of the more efficient cooling.  

For two other objects, 1ES\,0347-121 and 1ES\,1101-232, this scenario is not strictly necessary:
the UV and X-ray spectra can be fitted with a single synchrotron component (though not the WISE data in 1ES\,1101-232).
However, the resulting gamma-ray emission tends to slightly underestimate the frequency of the Compton peak with respect to the 
indication of the VHE spectra (i.e. below 1 TeV instead of above it, though the statistical uncertainty is large).
The Compton peak can be better reproduced  allowing again for a harder synchrotron spectrum below the X-ray band   
(compare for example curves {\it a} and {\it b} in Fig \ref{seds}).

The remaining two objects 1ES\,1218+304 and 1ES\,0414+009  are characterized by higher non-thermal emission at optical frequencies,
with respect to the other targets.  
In principle their SED can be fitted  using a single  electron distribution with three branches (i.e. two breaks).
However, as a result the Compton emission peaks at lower energies (Fig. \ref{seds}).
In 1ES\,1218+304, it gets underestimated with respect to the best-fit index of the EBL-corrected VERITAS spectrum
\citep[$\Gamma_{\rm VHE}\simeq 1.86\pm0.37$,][]{veritas1218},  by about one decade  
(below $\sim$200 GeV vs above 2 TeV).  Given the large statistical uncertainty, a 100-GeV peak energy is still compatible
with the data (since $\Gamma_{\rm VHE}\simeq2.23$ at 1$\sigma$), but it would imply that the source is not actually 
an extreme-Compton BL Lac but a more normal HBL.
A Compton peak above 2 TeV would again require a harder synchrotron spectrum which would not account for the observed UV flux.

1ES\,0414+009 is characterized by the lowest synchrotron-peak energy and softer X-ray spectrum in our sample. 
The non-thermal emission at optical frequencies swamps almost completely the host-galaxy flux, and indeed
all the WISE and UVOT data  appear to stay on a single power-law spectrum.
As explained in Sect. 3, this source has most likely changed its Compton peak properties between the VHE and \fermi-LAT epochs. 
We thus simply accounted for the \fermi-LAT emission as the gamma-ray counterpart of the \nustar~ pointings, 
obtaining a theoretical SED more in line with standard HBL (curve {\it a} in Fig. \ref{seds}).

However, if one attempts to model the hard VHE spectrum, 
the model parameters become similar to the ones for other sources (see e.g. curve {\it b} in Fig. \ref{seds}).
That is,  a much weaker magnetic field and consequently a strong deviation from
equipartition. Such a strong change of the magnetic field in the production site 
might be an important case for the verification of scenarios 
for the formation of gamma-ray production sites in BL LAC sources.

\begin{table*}
\centering
\begin{tabular}{lcccccccccccc}
\hline
\hline
Source  & $\gamma _{\rm 0}$ & $n_0$ & $\gamma _{\rm 1}$ & $\gamma _{\rm b}$& $\gamma _{\rm 2}$& $n_1$&$n_2$ &$B$ &$K$ &$R$ & $\delta$  & $U_{\rm e}/U_{\rm B}$   \\
\quad [1] & [2]  & [3] & [4] & [5] & [6] & [7] & [8]  & [9] & [10] & [11] & [12]  & [13]   \\
\hline
1ES 0229+200 a  & -  & -   &$100 $&$ 1.1\times 10^{6} $&$ 2\times 10^{7} $&$ 1.4 $&$ 3.35 $&$ 0.002 $&$ 6 $&$ 0.8 $& 50                          & $1.7\times10^{5}$  \\      
1ES 0229+200 b  & -  & -   &$2\times10^4 $&$ 1.5\times 10^{6} $&$ 2\times 10^{7} $&$ 2.0 $&$ 3.4 $&$ 0.002 $&$ 10^{3}$&$ 2.1 $& 50               & $2.0\times10^{4}$   \\	   
1ES 0347-121 a  & -  & -   &$100 $&$ 7.5\times 10^{5} $&$ 1.8\times 10^{7} $&$ 1.7 $&$ 3.8 $&$ 0.0015 $&$ 1.2\times 10^{2}$&$ 1.2 $& 60          & $1.5\times10^{5}$    \\      
1ES 0347-121 b  & -  & -   &$3\times10^3 $&$ 7.5\times 10^{5} $&$ 1.8\times 10^{7} $&$ 2.0 $&$ 3.8 $&$ 0.0015 $&$ 8\times 10^{2}$&$ 2.5 $& 60    & $3.4\times10^{4}$    \\      
1ES 0414+009 a  & 10 & 1.7 &$1\times10^4 $&$ 10^{5} $&$ 10^{6} $&$ 3.0 $&$ 4.6 $&$ 0.3 $&$ 8\times 10^{6}$&$ 2.1 $& 20                           &  0.5	    \\      
1ES 0414+009 b  & -  & -   &$3\times10^4 $&$ 5\times 10^{5} $&$ 3\times 10^{6} $&$ 2.0 $&$ 4.3 $&$ 0.0025 $&$ 1.6\times 10^{2}$&$ 6.5 $& 60      & $9.3\times10^{2}$    \\      
RGB J0710+591   & -  & -   &$100 $&$ 6\times 10^{5} $&$ 10^{7} $&$ 1.7 $&$ 3.8 $&$ 0.011 $&$ 1.2\times 10^{2}$&$ 0.92 $& 30                      & $2.7\times10^{3}$     \\      
1ES 1101-232 a  & -  & -   &$3.5\times 10^4$&$ 1.1\times 10^{6} $&$ 6\times 10^{6} $&$ 2.2 $&$ 4.75 $&$ 0.0035 $&$ 7.0\times 10^{3}$&$ 2.5 $& 60 & $2.4\times10^{3}$     \\      
1ES 1101-232 b  & -  & -   &$1.5\times 10^4$&$ 9.5\times 10^{5} $&$ 4\times 10^{6} $&$ 2.2 $&$ 4.75 $&$ 0.005 $&$ 2.4\times 10^{3}$&$ 3.8 $& 50  & $6.0\times10^{2}$    \\      
1ES 1218+304    &100 & 1.3 &$3\times 10^4 $&$ 10^{6} $&$ 4\times 10^{6} $&$ 2.85 $&$ 4.2 $&$ 0.0035 $&$ 1.2\times 10^{7}$&$ 3.5 $& 50            & $4.5\times10^{3}$     \\      
\hline
\hline
\end{tabular}
\vskip 0.4 true cm
\caption{
Input model parameters for the models in Fig. \ref{seds}.
[1]: Source.  
[2]: Minimum Lorentz factor (for the 3-power law model only).
[3]: Low energy slope of the electron energy distribution (3 power law model only).
[4], [5] and [6]: Minimum, break and maximum electron Lorentz factor.  
[7] and [8]: Slope of the electron energy distribution below and above $\gamma_b$. 
[9]: Magnetic field [G]. 
[10]: Normalization of the  differential electron distribution, in units of cm$^{-3}$. 
[11]: Radius of the emission zone in units of $10^{16}$ cm. 
[12]: Doppler factor. 
[13]: Ratio between the electrons energy density $U_{\rm e}$ and magnetic field $U_{\rm B}$.
}
\label{param}
\end{table*}

Considering the long integration times of the \fermi-LAT data (4 years) and that the VHE observations 
took place in general a few years before the \fermi~ observations, it is somewhat surprising that the LAT
and VHE data match so well. There seems to be a striking lack of long-term variability 
in the gamma-ray emission of these extreme sources.
  
A likely reason for this lack of large-amplitude variability is that blazars tend to be much less variable at energies below 
each SED peak than above it \citep[see e.g.][]{latvar},
the latter corresponding to radiation from freshly accelerated electrons.
The flux variations seen in these sources at UV and X-ray energies  have been historically much lower than 
in  more typical HBL  
like Mkn\,501 \citep[e.g.][]{pian98,tavecchio501,aliu16}
and Mkn\,421 \citep[e.g.][]{tramacere09,acciari14,kapanadze16,carnerero17}, 
a factor $2-5\times$ vs $100\times$ or more.
Besides, also normal HBL can spend years in low-activity states before undergoing huge flares 
\citep[e.g. PKS\,2155$-$304 in summer 2006,][]{2155ultrafast,2155chandranight,2155long17}.
   Variations of a factor a few in amplitude can remain  hidden in the statistical uncertainty and 
cross-normalization  of the gamma-ray datasets considered here.

\subsection{Cascades ?}
Another possibility is that the gamma-ray band is dominated 
by  secondary emission, arising from the interaction of 
primary gamma-rays with the diffuse EBL \citep[e.g.][]{jelley,nikishov,zdiarski88,halo94,coppi97}.
The produced electron-positron pairs generate secondary gamma-rays via IC scattering of 
the Cosmic Microwave Background (CMB),  reprocessing primary photons of energy $E_{\rm TeV}$ into lower-energy photons  
of energy  $E_{\gamma}\approx 0.63\, E_{\rm TeV}^2$ GeV \citep[see e.g.][]{tavecchio11}.
 
This interpretation however faces several difficulties for our targets.
The cascade scenario can be divided into two main cases:
{\it a)} both the LAT and the VHE data are dominated by secondaries;
{\it b)} only the LAT data are dominated by secondaries, while the VHE spectrum measures the primary blazar emission.

In the first case, a secondary emission peaking above 2-10 TeV  requires primaries above 50-100 TeV.
At such high energies, the mean free path for interactions with the cosmic infrared background is very small, 
of the order of 10 Mpc or less \citep{franceschini08,dominguez11}. 
At these distances from the AGN the magnetic field is expected to still be rather high ($10^{-9}-10^{-11}$ Gauss) 
--at the cluster or supercluster level \citep[e.g.][]{bruggen05,bruggen13}--  
and thus the pairs are quickly isotropized.  
These are the typical conditions for the formation of giant {\it pair halos} \citep{halo94,anant}.
However,  studies of the point spread function (PSF) in the H.E.S.S. and VERITAS  images of our targets
seem to exclude the presence of extended emission \citep{hessHalo,veritasHalo}. 
Furthermore, the spectrum should be softer than observed below  200 GeV
and the intrinsic primary flux is required to be $\sim 100\times$ higher than the flux from EBL-correction alone,
because of the re-isotropization \citep{anant}. 
This would increase dramatically the energy requirements of the source, both in luminosity and peak energy
of the primary emission, making the problem of explaining the emitted SED even worse. 

The second possibility is that the LAT photons arise from the reprocessing of  primary VHE gamma-rays. 
The intrinsic blazar emission would peak in the $\sim2-20$ TeV range, as in the leptonic SSC scenario. 
Only one generation of pairs is produced,      
reprocessing all the absorbed power in the LAT band.  The primary emission beam is thus broadened in solid angle 
by the intergalactic magnetic field (IGMF) acting on the first-generation pairs.
The percentage of the absorbed primary flux reaching the observer within the LAT PSF
--and thus adding to the primary flux--  depends then  on the intensity and filling factor of the IGMF along the line of sight  
(assuming the EBL isotropic).  Indeed the \fermi-LAT data of our targets have already been used to put 
important constraints  on the IGMF, at the level of $B\gtrsim 10^{-15} - 10^{-18}$ depending on the 
assumed duration of source activity
\citep{neronov10,tavecchio11,taylor11,finke15}.

This possibility requires on the one hand a small IGMF (below or close to the derived lower limits) 
and a very small filling factor, 
for the secondary emission to be dominant and within the LAT PSF \citep{latpsf}. 
Both conditions are not granted given the large scale structure 
of the local Universe \citep[e.g.][and references therein]{costamante13}.
On the other hand, the total amount of absorbed power and the primary spectrum is known, from the observed VHE data 
and EBL density \citep{nature_ebl,lat_ebl}.
This power must emerge in the LAT band on top of any underlying primary GeV emission.
The latter  therefore must 
be kept well below   
the extrapolation of the VHE spectrum into the GeV range, by one to two orders of magnitude.
In order not to overproduce the observed LAT data, the primary radiation --and its corresponding electron distribution-- 
must be very narrowly peaked around 2-10 TeV. A narrow electron distribution cannot in general reproduce the full broad-band
synchrotron emission in the SED \citep{lefa11}.  Other electrons could be responsible for different parts of the synchrotron hump, 
but their SSC emission would again fill the LAT band with primary radiation, 
unless their SSC flux is suppressed by assuming high magnetic fields in their emitting region.
This possibility seems thus less likely due to the extreme fine-tuning and ad-hoc conditions required.

We conclude that, though some contribution from secondary radiation cannot be excluded,
it should not be the dominant component of the observed gamma-ray flux.

\section{Discussion and Conclusions}
The combined \nustar~  and \swift~ observations  provide for the first time three important information for these objects: 
1) the precise location of the synchrotron peak in the SED, also for the hardest objects;
2) the relation of the UV flux with respect to the X-ray spectrum;
and 3) the absence of a significant hardening of the emission towards higher X-ray energies.
The latter result goes against the idea that the hard TeV spectra are produced by an additional electron population,
emitting by synchrotron in the hard X-ray band.
Their emission is constrained to be well below the observed flux (i.e. implying a high Compton dominance) 
or at energies much above 100 keV.

Using archival \fermi-LAT and VHE observations, we built the best sampled SED so far for these objects, 
and tested the one-zone SSC scenario.
A leptonic SSC model is able to reproduce the extreme properties of both peaks in the SED quite well, 
from X-ray  up to TeV energies,  but at the cost of  i)  extreme acceleration and very low radiative efficiency,
with conditions  heavily out of equipartition   (by 3 to 5 orders of magnitude);
and ii)  dropping the requirement to match the simultaneous  UV data,  which then should 
belong to a different zone or emission component,  possibly the same as the far-IR (WISE) data.

This scenario is corroborated by direct evidence in the X-ray data of   
1ES\,0229+200 and RGB\,J0710+591.
Their UV flux is in excess of 
the extrapolation of the soft X-ray spectrum to lower energies.
The model can be made to reproduce well 
either the UV data (over-estimating the soft-X spectrum) 
or the soft-X spectrum (under-estimating the UV flux), but not both. 
In the other sources this scenario is not strictly necessary but 
becomes preferable
in order to fully reproduce a Compton peak at multi-TeV energies.

The discrepancy between particle and magnetic energy density is dramatic.
Considering a more accurate geometry in the number density of synchrotron photons
inside a region of homogeneous emissivity \citep[see][]{atoyan96} can bring the conditions 
a factor 3-4 closer to equipartition, but cannot account for orders of magnitude.
Remarkably, this discrepancy would not widen significantly 
in presence of hot protons in the jet, instead of the more common cold assumption.
The reason is that the average electron energy in these sources is higher than the rest mass of the proton.

Conditions so far away from equipartition are even more puzzling  since not limited to a flaring episode:
the extreme nature of the SED in these BL Lacs seem to last for years.
If the leptonic scenario is correct, there must be a mechanism which keep the conditions
in the dissipation region  persistently out of equipartition.
The specific case of 1ES\,0414+009 shows, however, that some sources could possibly switch closer to equipartition
after some years.

In our modeling, the size of the emitting region is of the order of $R\sim 10^{16}$ cm 
with high Doppler factors of 30-60 (see Table \ref{param}). These values can accomodate variability
on a daily timescale like the one shown by 1ES\,1218+304 \citep{veritas1218var}.
In principle, it would be possible to have smaller Doppler factors with larger sizes of the 
emitting region (for example, in 1ES\,0229+200, $\delta=10$ with $R\sim 10^{17}$ cm).
However, this solution cannot accomodate variability much faster than a week,
and it does not help in bringing the conditions much closer to equipartition 
(in our example, $U_{\rm e}/U_{\rm B}\sim 1.2\times10^{5}$).

These hard-TeV BL Lacs represent the extreme case of the more general problem of magnetization in BL Lacs, 
for which one-zone models imply particle energy and jet  kinetic power largely exceeding the magnetic power 
\citep[see e.g.][]{tavecchio16}.
In these extreme-Compton objects even the assumption of a structured jet 
--namely a fast spine surrounded by a slower layer--
does not help in reaching equipartition. 
If the layer synchrotron emission is sufficiently broad-banded,
the additional  energy density in soft photons provided by the layer to the fast-spine electrons 
does allow for a larger magnetic field 
and higher IC luminosity, but 
would generate more efficient cooling of the TeV electrons, preventing a hard spectrum at TeV energies.
A spine-layer scenario can thus give solutions close to equipartition  
for ``standard" HBL with a soft TeV spectrum \citep[e.g.][]{tavecchio16}, 
but not for these hard-TeV BL Lacs.

The high values of the ratio between the energy density of particles $U_{\rm e}$ and magnetic field $U_{\rm B}$
($U_{\rm e}/U_{\rm B}\sim 10^3-10^5$)  seems also to exclude magnetic reconnection as possible mechanism 
for accelerating electrons. At present, relativistic reconnection 2D models predict an upper limit of the order of 
$U_{\rm e}/U_{\rm B}\sim 3$ in the dissipation region \citep{sironi15}.
The presence of a guiding magnetic field should reduce this ratio even further ($U_{\rm e}/U_{\rm B}\sim 1$).

The \nustar~  and \swift~  observations of our targets are consistent with the blazar sequence of SED peak frequencies
\citep[e.g.][]{sequence2},
showing a  correlation between synchrotron and Compton peak energies.
Namely, the highest synchrotron peak frequency and hardest X-ray spectrum are reached in the object 
with the highest Compton peak energy (1ES\,0229$+$200),
and vice-versa (1ES\,0414+009), with the other four targets clustering around few keV and few TeV for the two SED peaks. 

This behaviour is different from standard HBL during strong flares.
Standard BL Lacs  can show  extreme-synchrotron properties in the X-ray band during flares,
like for example Mkn\,501 in 1997 \citep{pian98,hegra501spec,501model} and 2009 \citep{veritas501,paneque16};
or  1ES\,1959$+$650 in 2002 \citep{kraw1959} and 2016 \citep{atel1959x,atel1959gamma}.
During these flares the synchrotron X-ray emission hardens significantly, shifting the synchrotron peak from below 0.1 keV 
to 5-10 keV up to 100 keV or more. However, that does not bring a corresponding shift of the 
Compton peak towards extreme values.
The intrinsic VHE emission in those objects remains always steep ($\Gamma_{\rm VHE}>2$), 
locating the gamma-ray peak in the SED 
below or around few hundreds GeV \citep[possibly close to $\sim1$ TeV only in Mkn\,501, in a particular episode, e.g.][]{paneque16}. 
This might indicate the presence of stronger magnetic fields or generally more efficient cooling
during strong dissipative events,
as well as the result of a single zone dominating the overall emission from optical to TeV energies.

Given their puzzling emission properties,  it is important to find and study more
objects of this new type \citep[see e.g.][]{bonnoli15}.
The extreme-Compton nature is clearly revealed only through VHE observations, by measuring hard TeV spectra
after correction for EBL absorption, irrespective of the X-ray or GeV spectra.
This requires VHE telescopes  with the largest possible collection area and sensitivity 
in the multi-TeV range, given the typical fluxes of at most few percent of the Crab flux. 
The upcoming ASTRI \citep{astri} and CTA \citep{cta} air-\C arrays  will provide significant progress in this respect.

\section*{Acknowledgements}
We acknowledge financial support from the CaRiPLo Foundation and
the regional Government of Lombardia with ERC for the project ID 2014-1980 
``Science and technology at the frontiers of gamma-ray astronomy
with imaging atmospheric Cherenkov Telescopes'', and from the 
agreement ASI-INAF I/037/12/0.
We thank Gino Tosti for useful discussions.
L. Costamante thanks the Japan Aerospace Exploration Agency (JAXA) for the
hospitality and financial support during his visit. 
This work made use of data from the \nustar~ mission, a project led by
the California Institute of Technology, managed by the Jet Propulsion
Laboratory, and funded by the National Aeronautics and Space
Administration. We thank the \nustar~ Operations, Software and Calibration teams 
for support with the execution and analysis of these observations.
This research has made use of the NuSTAR Data Analysis Software (NuSTARDAS) jointly developed 
by the ASI Space Science Data Center (SSDC, Italy) and the California Institute of Technology
(Caltech, USA). 
We are grateful to Niel Gehrels and the \swift~ team for executing and making possible the
simultanous observations. 
This research has made use of the XRT Data Analysis Software (XRTDAS)
developed under the responsibility of the ASI Space Science Data Center (SSDC), Italy.
Part of this work is based on archival data, software or on-line services provided by SSDC.
This research has made use of the NASA/IPAC Extragalactic Database (NED) which is operated by the Jet Propulsion Laboratory, 
California Institute of Technology, under contract with the National Aeronautics and Space Administration.

We dedicate this paper to Niel Gehrels, who passed away prematurely.  
His personality and leadership was an essential part of the \swift~ success.
He will always have a special place in our memories.

\bibliographystyle{mnras}
\bibliography{biblioteca} 

\bsp	
\label{lastpage}
\end{document}